# 2DMatPedia: An open computational database of two-dimensional materials from top-down and bottom-up approaches

**Jun Zhou[1], Lei Shen[2], Miguel Dias Costa[3], Kristin A. Persson[4,5], Shyue Ping Ong[6], Patrick Huck[4], Yunhao Lu[7], Xiaoyang Ma[1] & Yuan Ping Feng[1,3]**

[1]Department of Physics, National University of Singapore, Singapore 117411, Singapore
[2]Department of Mechanical Engineering, National University of Singapore, Singapore 117575, Singapore
[3]Centre for Advanced Two-dimensional Materials, National University of Singapore, Singapore 117546, Singapore.
[4]Department of Materials Science and Engineering, University of California Berkeley, California 94720, USA
[5]Lawrence Berkeley National Laboratory, Berkeley, California 94720, USA
[6]Department of NanoEngineering, University of California, San Diego, 9500 Gilman Drive, La Jolla, California 92093, USA
[7]State Key Laboratory of Silicon Materials, School of Materials Science and Engineering, Zhejiang University, Hangzhou 310027, China
corresponding author(s): Yuan Ping Feng (phyfyp@nus.edu.sg)

Two-dimensional (2D) materials have been a hot research topic in the last decade, due to novel fundamental physics in the reduced dimension and appealing applications. Systematic discovery of functional 2D materials has been the focus of many studies. Here, we present a large dataset of 2D materials, with more than 6,000 monolayer structures, obtained from both top-down and bottom-up discovery procedures. First, we screened all bulk materials in the database of Materials Project for layered structures by a topology-based algorithm, and theoretically exfoliate them into monolayers. Then, we generated new 2D materials by chemical substitution of elements in known 2D materials by others from the same group in the periodic table. The structural, electronic and energetic properties of these 2D materials are consistently calculated, to provide a starting point for further material screening, data mining, data analysis and artificial intelligence applications. We present the details of computational methodology, data record and technical validation of our publicly available data (http://www.2dmatpedia.org/).

## Background & Summary

Atomically thin two-dimensional (2D) materials have attracted tremendous research interest for both novel fundamental physics and extremely appealing applications. For example, new emerging physics such as half-integer quantum Hall effect[1], Klein tunnelling[2], valley Hall effect[3] and topological superconductivity[4, 5] have been reported in 2D materials. 2D structures are naturally beneficial for performance of various types of devices, such as large surface-to-volume ratio for high sensing sensitivity and catalysis efficiency[6, 7], reduced size for immunity against short channel effects[8] and flexibility for wearable devices[9, 10], to name only a few. Furthermore, the Van-der-Waals stacking of homo/hetero 2D materials provides another degree of freedom to tune the desired properties of the system[11, 12]. However, only dozens of 2D materials have been experimentally synthesised. New 2D materials with novel properties are needed to meet the demand of ever growing technological applications.

The traditional material discovery is mainly based on trial-and-error experiments which is time consuming and resource intensive. To accelerate development and deployment of novel advanced materials, the US White House launched the "Materials Genome Initiative" in 2011[13]. This approach integrates high throughput computation, data analytics together with experimental research and represents a new paradigm for materials discovery. The data-driven material discovery can significantly reduce the cost from many long iterations of trial-and-error experiments by providing the most promising candidates from high-throughput computations. This approach is also more flexible as different screenings can be conducted to target materials with specific properties for different applications. In this spirit, large repositories with millions of computed bulk material entries have been developed such as the Materials Project (MP)[14], the Open Quantum Materials Database (OQMD)[15, 16], the Automatic Flow for Materials Discovery (AFLOWLIB)[17], and the Novel Materials Discovery (NOMAD) Repository [18], thanks to the development of computing power and significant advancements of the accuracy of first-principles calculations.

Databases specific for 2D materials have also emerged rapidly from small repositories with tens of entries within some specific structure prototypes in earlier works [19-23] to more recent systematic data mining of the Materials Project, Inorganic Crystal Structure Database (ICSD)[24] and the Crystallographic Open Database (COD) [25], yielding thousands of monolayers[26-29]. Nevertheless, most of the current databases for 2D materials have been obtained mainly using the top-down approach, that is, by theoretically exfoliating monolayers from layered bulk materials. And some criteria on the stability and exfoliation energy were applied to predict potentially exfoliatable stable monolayers[26, 28]. This method allows a systematic screening of 3D materials for layered structures. However, the top-down approach is limited in at least three cases. i) Some 2D materials, such as silicene[30], do not have corresponding layered bulk forms. ii) A 2D material, such as $MoS_2$[31], can have a few stable polymorphs but only one of them has a corresponding bulk form. iii) Existing 3D material databases are not comprehensive enough and some known materials are not captured in these databases. For example, the bulk form of the seminal 2D ferromagnetic $CrI_3$ is unfortunately absent in Materials Project[14]. On the other side, chemical substitution has been shown to be an efficient way to predict new 2D materials[22, 32]. Furthermore, 2D materials with relatively high exfoliation energies can be grown by other methods than mechanical exfoliation, and metastable 2D materials can be stabilized by means such as supported on a substrate[30, 33].

Here, we present an alternate database, 2D Materials Encyclopedia (2DMatPedia), with monolayers obtained from both the top-down and the bottom-up approaches. Unbiased by energetic stability, we screen all the possible layered bulk materials from the Materials Project by a topology-based algorithm, and theoretically exfoliate them into monolayers. Then, we systematically generate new 2D materials by elemental substitution of the unary and binary compounds obtained from the top-down approach. This combined top-down and bottom-up approach has generated more than 6,000 2D structures. Their structural, electronic and energetic properties have been obtained by high-throughput calculations. This large and consistent 2D material dataset could serve as a good starting point for further materials screening, data mining and training of machine learning (ML) models. The whole database is publicly available at http://www.2dmatpedia.org/.

## Methods

The methodology applied in this work mainly includes two parts: discovery processes to generate 2D structures and high-throughput density functional calculations of properties of the 2D materials. The overall workflow is shown in Fig. 1 and the details of each step are explained below.

### Discovery processes

We use both a top-down approach, in which materials from the inorganic bulk crystals in the Materials Project are screened for layered structures which are then theoretically exfoliated to 2D monolayers, and a bottom-up approach, in which elemental substitution is systematically applied to the unary and binary 2D materials obtained from top-down approach. It is noted the "discovery process" here only indicates how a 2D material is generated in this work, which is not necessarily related to its experimental synthesis method. For example, 2D $CrI_3$ as mentioned above is generated by a "bottom-up" approach here but was exfoliated experimentally[34].

### Top-down approach

A geometry-based algorithm[26, 28] was used to identify layered structures among these compounds by the following steps:

1. Standard conventional unit cell was used for all the compounds.
2. Check whether two atoms in the unit cell are bonded or not. Use the sum of the covalent radii [35] of two elements with a small tolerance as a threshold. If the distance of two atoms is smaller than the threshold, they are considered boned. Otherwise, they are not bonded.
3. Atoms that are bonded together form a cluster. The number of clusters in the unit cell is counted.
4. From the unit cell, a 3×3×3 supercell is generated and the number of clusters in the supercell is counted again. If the number of clusters in the supercell is three times of that in the unit cell, the structure was tagged as layered.
5. A set of tolerances (0.0, 0.05, 0.1, 0.15, 0.2, 0.25, 0.3, 0.35, and 0.4) was used and only the structures identified as layered by at least two tolerances are kept.
6. 2D materials were theoretically exfoliated by extracting one cluster in the standard conventional unit cell of the layered structures screened in the above step. A larger than 20 Å vacuum along *c* axis was imposed to minimize the interactions of image slabs by periodic condition.

### Bottom-up approach

1. All the elements of the periodic table (from H to Bi) are categorized in different groups according to their column number. Radioactive elements, lanthanide and actinide with *f* electrons are excluded except La which is included in group 3 elements.
2. Systematically replace each element in a known 2D material by one other element in the same category. For instance, 24 new materials can be generated from BN by replacing B with [B, Al, Ga, In, Tl] and N with [N, P, As, Sb, Bi] systematically.

## Workflow

As shown in Fig. 1, we started from the > 80000 inorganic compounds in Materials Project database. In the initial stage, we focus on elemental, binary, ternary and quaternary compounds and also ignore structures with more than 40 atoms in the primitive unit cell. The top-down approach discussed above was applied to screen the database for layered structures and generate 2D materials. Structure matching tools from pymatgen[36] were used to remove duplicates in the exfoliated 2D materials. High-throughput calculations, adopting the same standards set by the Materials project, were carried out to optimize the structures, and perform static, density of states (DOS) and band structure calculations for these 2D materials. The calculated properties are stored in 2DMatPedia. The unary and binary 2D materials obtained from the top-down approach were then used as initial structures for elemental substitution. Structure matching was applied again to these 2D materials obtained through this bottom-up approach to ensure that only unique structures are included for further high-throughput density functional theory (DFT) calculations.

## High-throughput calculations

The standard workflow[37, 38] developed by Materials Project was used to perform high-throughput calculations for all the layered bulk and 2D materials generated in the above processes. The DFT calculations were performed using the Vienna Ab initio Simulation Package (VASP) with the Perdew-Burke-Ernzerhof (PBE) approximation for the exchange-correlation functional and the frozen-core all-electron projector-augmented wave (PAW) method for the electron-ion interaction[39, 40]. The cutoff energy for the plane wave expansion of electron wavefunction was set to 520 eV. The interlayer dispersion interaction in layered materials is included via the dispersion-corrected vdW-optB88 exchange-correlation functional[41-44]. This functional has been shown to reproduce reasonably well the results from the more accurate but computationally demanding random-phase approximation (RPA) calculations [26, 45].

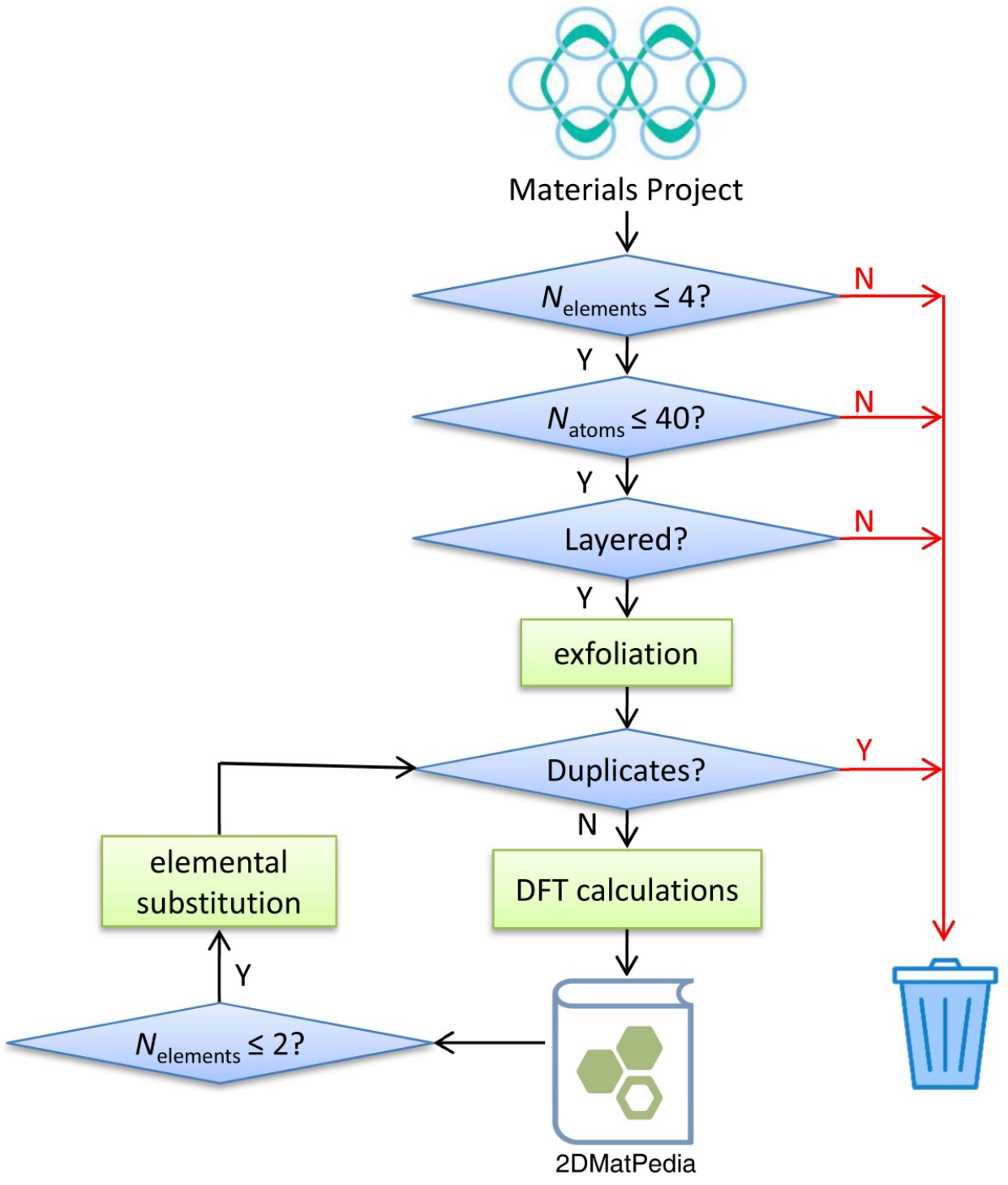

Figure 1. The workflow of producing data of 2D materials.

**Structure optimization**

For structure optimization of 2D materials, both cell shape and volume were adjusted (ISIF = 3) while the *c* axis was kept fixed. The energy difference for ionic convergence is set to $1.0\times10^{-4}$ eV. The dispersion-corrected vdW-optB88 exchange-correlation functional was applied for structure relaxation.

**DOS and band structure calculations**

A static run with a uniform ($\Gamma$-centered) k-point grid for the relaxed structure is performed first to generate the charge density. Non-self-consistent runs based on the charge density with an energy range from $E_F$ (Fermi level)-10 eV to $E_F$+10 eV with 2000 intervals were used for DOS simulation. The band structure computation uses line-mode k-point grid along the high symmetry points of the 2D Brillouin zone [26, 46]. Van der Waals correction is not applied in these calculations.

**Exfoliation Energy and Decomposition energy**

The exfoliation energy, the average energy per atom required to remove a layer from a layered bulk material, is calculated from by $E_{exf} = E_{2D}$-$E_{bulk}$, where $E_{2D}$ and $E_{bulk}$ are the total energy per atom of the 2D and its layered bulk counterpart, respectively. To obtain the total energy, similar structure optimization and static calculations are performed for both the 2D and bulk materials. The total energy of the static run is used to calculate the exfoliation energy. The dispersion-corrected vdW-optB88 exchange-correlation functional is applied to all the runs in the exfoliation energy calculations.

However, not every 2D material will have a unique bulk counterpart. To ensure a meaningful comparison of exfoliation energies of various 2D materials, the following assumptions were made. If a 2D material has only one corresponding layered bulk in the Materials Project, this bulk material is tagged as its layered bulk counterpart. If there are multiple layered bulk structures in MP corresponding for a given 2D material, only the one with the lowest energy is tagged as its layered bulk counterpart. For 2D materials obtained through the bottom-up approach, there is often no existing corresponding layered bulk material. In such a case, we also construct a "layered bulk" counterpart by stacking the monolayers following the same sequence of the initial structure (i.e., BN in the example given earlier) is stacked in its bulk counterpart. It is noted that for a few special cases like Silicene, the "exfoliation energy" is provided only for reference. The energetic stability of such a 2D material is better estimated by the decomposition energy as discussed below.

The decomposition energy is defined as the energy required to separate a compound into its components. Here, we applied a modified version of the energy_above_hull, and define the decomposition energy as the energy required (per atom) to decompose a given material into the set of most stable materials at this chemical composition[47]. This definition of decomposition energy excludes the corresponding layered bulk of the given 2D material from its competing phases. The energy difference between a 2D material and its layered bulk is given by the exfoliation energy.

The competing entries are queried from Materials Project, which are calculated without the Van der Waals correction. In order to make direct comparison, we also calculate the total energies of 2D materials without the Van der Waals correction. This energy and the energies of all entries belonging to the same chemical system in Materials Project are used to generate a phase diagram. The decomposition energy is obtained from the difference between the energy of the 2D material and those of its competing phases.

**Code availability**

The code used in this work relies heavily on the tools developed by Materials Project (https://github.com/materialsproject;https://github.com/hackingmaterials). The versions used in this work are pymatgen/4.7.3 and atomate/0.5.1 for all the calculations.

## Data Records

The web interface of 2DMatPedia can be found at http://www.2dmatpedia.org/. A JASON file is available at http://www.2dmatpedia.org/download. Table 1 shows the key variables of this materials collection, including name, data type and a short description. The 'material_id' is the identifier of the each unique 2D material in the dataset. The 'relative_id' provides easy link to the material from which this 2D material is obtained. The 'discovery_process' shows how the 2D material is generated ("top-down" or "bottom-up"). The 'structure' is the relaxed structure in a dictionary format. It can be easily transferred to different format by pymatgen. Other information that can be deduced from a structure, such as chemical formula ('formula'), number of elements in the structure ('nelements'), list of elements ('elements'), ' spacegroup', ' point_group', are also listed for easy query. The documented electronic, energetic and magnetic properties are 'bandgap', 'is_gap_direct', 'is_metal', 'energy_per_atom', 'energy_vdw_per_atom', 'exfoliation_energy_per_atom', 'decomposition_energy' and 'total_magnetization''.

| Key | Datatype | Description |
|---|---|---|
| material_id | string | IDs for entries in the 2Dmatpedia |
| relative_id | string | IDs for where a 2D material is obtained from |
| discovery_process | string | How a 2D materials is generated |
| structure | dictionary | Relaxed crystal structure represented in dictionary |
| formula | string | Chemical formula |
| nelements | string | Number of elements in this material |
| elements | list | List of elements in this material |
| spacegroup | string | Space group number defined by The International Union of Crystallography |
| point_group | string | Point group in Hermann-Mauguin notation |
| bandgap | float | Energy band gap of this material |
| is_gap_direct | Boolean | Is the material a direct gap |
| is_metal | Boolean | Is the material metallic |
| energy_per_atom | float | Energy per atom in eV without vdW correction |
| energy_vdw_per_atom | float | Energy per atom in eV with vdW correction |
| exfoliation_energy_per_atom | float | Exfoliation energy of the 2D material in eV/atom |
| decomposition_energy_per_atom | float | Decomposition energy of the 2D material in eV/atom |
| total_magnetization | float | Total magnetic moment in $\mu_B$ |

**Table 1.** JSON keys for metadata and their descriptions.

## Technical Validation

We have performed the following validations for our results in 2DMatPedia. First, we benchmarked the structures and energies against data in an existing 2D material database, JARVIS, which applied well-converged energy cut-off and k-mesh density. Then we analyse the calculated decomposition energies for 59 experimentally grown 2D materials.

### Convergence

We apply the standard workflow designed by the Materials project for two reasons. (1) The parameters for energy cutoff, k-point mesh, and threshold of energy convergence are extensively tested by the Materials project (https://materialsproject.org/docs/calculations). (2) It makes our results compatible with the entries in the Materials project, and we could make direct comparisons to get useful information such as decomposition energy. To evaluate the performance of these standard workflow in predicting structures and energies of 2D materials, we compare the calculated lattice parameters and exfoliation energies with those in the JARVIS database (https://figshare.com/articles/jdft_2d-7-7-2018_json/6815705).

As shown in Fig. 2, the calculations in this work reproduced very well the results in JARVIS and the differences in lattice constants and exfoliation energies between 2DMatPedia and JARVIS are very small. The mean error and standard deviation are 0.0090 Å and 0.034 for lattice parameter *a*, 0.010 Å and 0.062 for lattice parameter *b*, -1.13 meV/atom and 5.64 for exfoliation energy, respectively. Among the 383 data collected, 93% of them have an exfoliation energy difference smaller than 10 meV/atom.

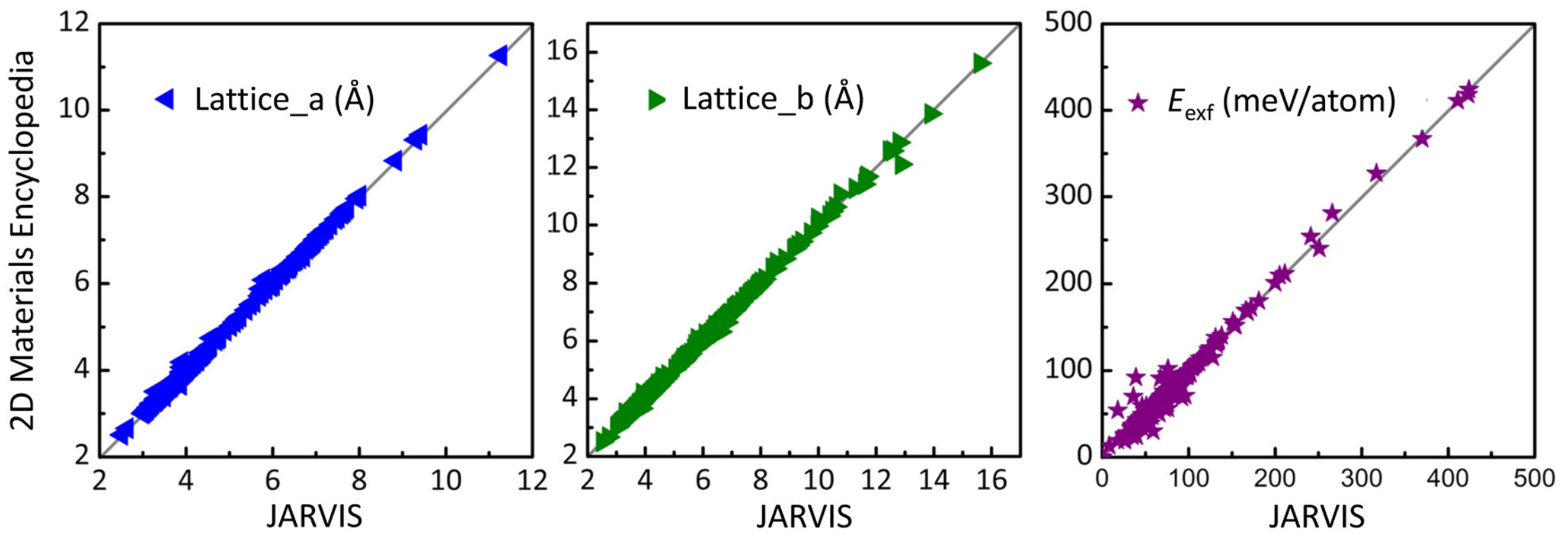

**Figure 2.** Structural and energetic comparison of the results from JARVIS and in this work. (left) and (center) lattice constants of *a* and *b*, respectively. (right) exfoliation energy.

### Decomposition energy

Figure 3 shows the histogram of decomposition energy for 2D materials obtained from both top-down and bottom-up approaches in 2DMatPedia. Of the 2,884 2D materials obtained via the top-down, more than 1,500 could be energetically stable, with a low decomposition energy of less than 100 meV/atom, while for those obtained through the bottom-up design, around 900 out of 2,927 are considered stable. The overall distribution shows that the number of compounds obtained from the top-down approach drops very fast with the decomposition energy, and almost vanishes at 1600 meV/atom, while those obtained from bottom-up approach decreases slowly and has significant number of compounds even at 1800 eV/atom. This can be understood by that top-down approach starts from the existing bulk materials while the bottom-up compounds are generated by artificial elemental substitution. Nevertheless, we keep these less stable 2D structures in our database, as such structures may have good functionalities and they could be stabilized by other means such as by a substrate. The large number of relative stable 2D compounds obtained from the bottom-up approach shows its important complimentary role to top-down method to generate a more comprehensive 2D materials dataset.

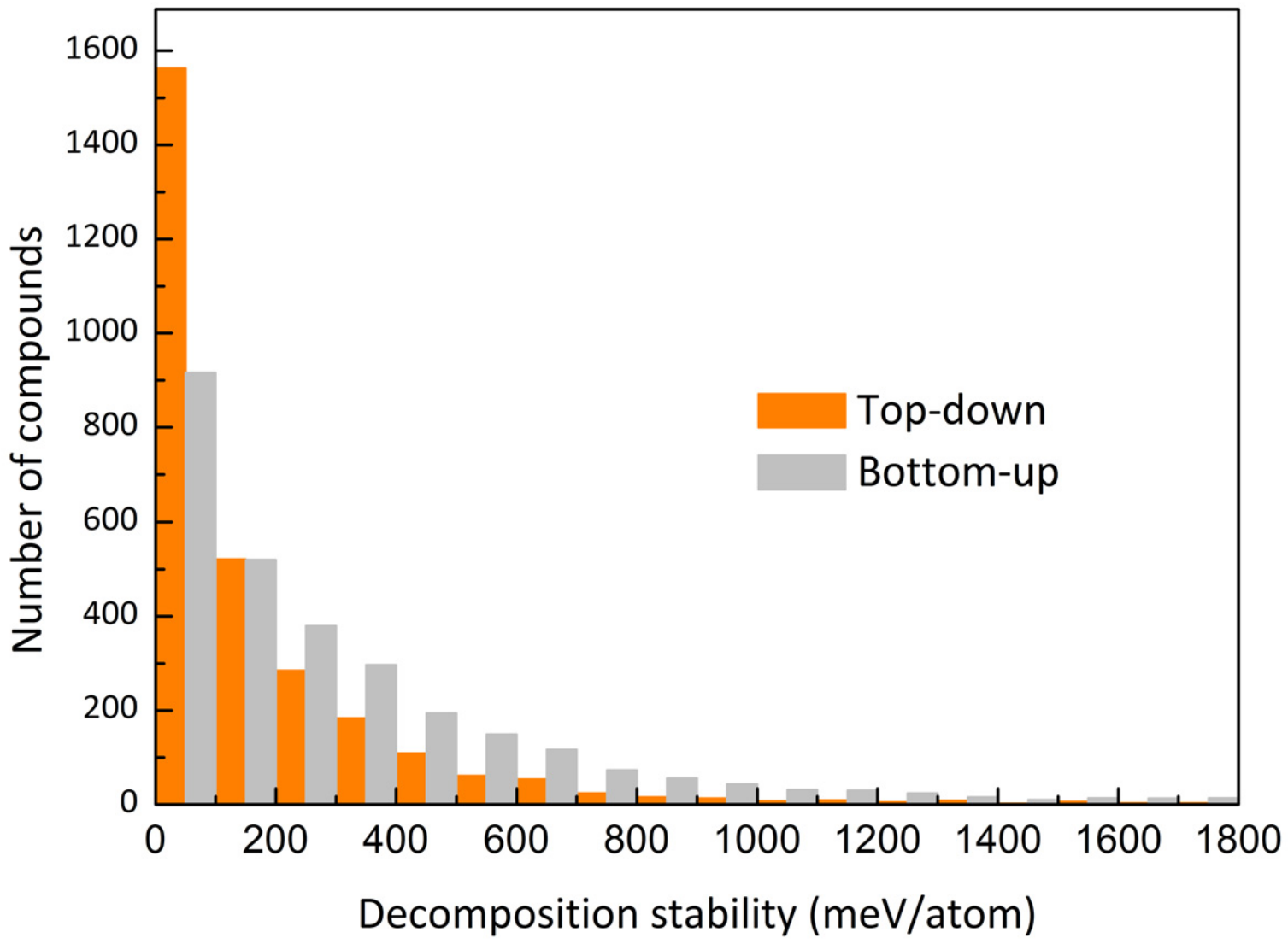

**Figure 3.** Histogram of the decomposition energy for both the top-down and bottom-up 2D materials in this work.

Figure 4 shows the calculated decomposition energy in this work for the 59 experimentally grown 2D materials. It is noted that 53 out of them have decomposition energies of less than 100 meV/atom and 39 compounds within 10 meV/atom. The 4 compounds with a relatively high value (> 150 meV/atom), that is, 2D $WO_3$, T'-$MoS_2$, Ge (germanene), Si (silicene), are known to be metastable in free standing form but have been synthesized on metal substrates or by metal atom intercalation.[30, 48-50] $Ca_2N$ has a high exfoliation energy (260 meV/atom in this work which is in good agreement with previous work[51]) but has been exfoliated into 2D nanosheets by liquid exfoliation and suspend in organic solvents. It may represent the highest possible decomposition energy of a monolayer that can be exfoliated to free standing films.

In Fig. 4, compounds discovered by the top-down approach are indicated by blue open squares and those obtained by the bottom-up approach are shown in red solid squares. It is obvious that the top-down approach alone failed to capture some of the experimental grown compounds. And the bottom-up generated 2D materials such Silicene, 1T'-$MoS_2$ and $CrI_3$ demonstrate its complimentary role to the top-down approach in the discovery of 2D materials.

Overall, our calculations have yielded reasonable values of decomposition energy for 2D materials, and this quantity can be used as a guide for experimentally synthesis of the 2D materials.

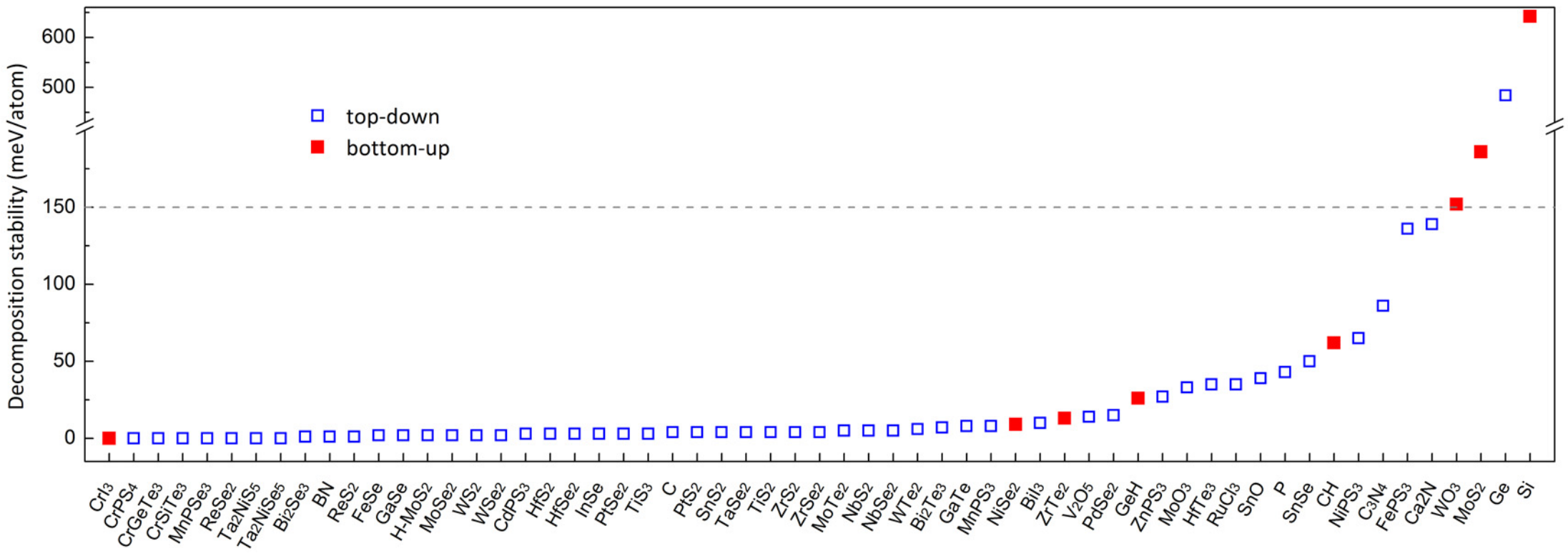

**Figure 4.** Decomposition energy calculated in this work for 59 experimentally grown 2D materials [30, 33, 34, 48-94]. The blue open squares are 2D materials generated in a top-down approach while the red solid squares are the ones from the bottom-up approach in this work.

## Usage Notes

The dataset in this work presents a large collection of > 6000 2D materials with consistently calculated structural, energetic and electronic properties. Currently, the database includes unary and binary bottom-up 2D compounds by systematically elemental substitution. Such a method cannot be directly applied to ternary and quaternary compounds for their huge phase space. We are developing machine learning algorithms to perform a preliminary screening of the large phase space of the elemental substituted ternary and quaternary 2D materials to exclude the extremely unstable compounds and performing high-throughput calculations for the left. And the database is also growing with new 2D materials by compounds from literature and 2D alloys. The user will be able to scan the entire database to screen for 2D materials with new functionalities, data mining, data analysis, and artificial intelligence applications.

## Acknowledgements

We acknowledge funding support of Singapore Ministry of Education Academic Research Fund Tier 1. We also thank the Centre for Advanced Two-dimensional Materials at National University of Singapore and the Singapore National Supercomputing Centre for providing computing resource.

## References

1. Zhang Y, Tan Y-W, Stormer HL, Kim P. Experimental observation of the quantum Hall effect and Berry's phase in graphene. *Nature* **438**, 201 (2005).
2. Katsnelson MI, Novoselov KS, Geim AK. Chiral tunnelling and the Klein paradox in graphene. *Nat. Phys.* **2**, 620 (2006).
3. Schaibley JR*, et al.* Valleytronics in 2D materials. *Nat. Rev. Mater,* **1**, 16055 (2016).
4. Wang J, Xu Y, Zhang S-C. Two-dimensional time-reversal-invariant topological superconductivity in a doped quantum spin-Hall insulator. *Phys. Rev. B* **90**, 054503 (2014).
5. Cao Y*, et al.* Unconventional superconductivity in magic-angle graphene superlattices. *Nature* **556**, 43 (2018).
6. Luo B, Liu G, Wang L. Recent advances in 2D materials for photocatalysis. *Nanoscale* **8**, 6904-6920 (2016).

7. Schedin F, *et al.* Detection of individual gas molecules adsorbed on graphene. *Nat. Mater.* **6**, 652 (2007).
8. Chhowalla M, Jena D, Zhang H. Two-dimensional semiconductors for transistors. *Nat. Rev. Mater,* **1**, 16052 (2016).
9. Akinwande D, Petrone N, Hone J. Two-dimensional flexible nanoelectronics. *Nat. Commun.* **5**, 5678 (2014).
10. Gao L. Flexible Device Applications of 2D Semiconductors. *Small* **13**, 1603994 (2017).
11. Geim AK, Grigorieva IV. Van der Waals heterostructures. *Nature* **499**, 419 (2013).
12. Mak KF, Lee C, Hone J, Shan J, Heinz TF. Atomically Thin $MoS_2$: A New Direct-Gap Semiconductor. *Phys. Rev. Lett.* **105**, 136805 (2010).
13. Material Genome Initiative. https://www.mgi.gov/
14. Jain A, *et al.* Commentary: The Materials Project: A materials genome approach to accelerating materials innovation. *APL Mater.* **1**, 011002 (2013).
15. Kirklin S, *et al.* The Open Quantum Materials Database (OQMD): assessing the accuracy of DFT formation energies. *Npj Comput. Mater.* **1**, 15010 (2015).
16. Saal JE, Kirklin S, Aykol M, Meredig B, Wolverton C. Materials Design and Discovery with High-Throughput Density Functional Theory: The Open Quantum Materials Database (OQMD). *JOM* **65**, 1501-1509 (2013).
17. Curtarolo S, *et al.* AFLOW: An automatic framework for high-throughput materials discovery. *Comput. Mater. Sci.* **58**, 218-226 (2012).
18. The NOMAD Laboratory. http://nomad-repository.eu; https://nomad-coe.eu
19. Miró P, Audiffred M, Heine T. An atlas of two-dimensional materials. *Chem. Soc. Rev.* **43**, 6537-6554 (2014).
20. Şahin H, *et al.* Monolayer honeycomb structures of group-IV elements and III-V binary compounds: First-principles calculations. *Phys. Rev. B* **80**, 155453 (2009).
21. Lebègue S, Björkman T, Klintenberg M, Nieminen RM, Eriksson O. Two-Dimensional Materials from Data Filtering and Ab Initio Calculations. *Phys. Rev. X* **3**, 031002 (2013).
22. Ataca C, Şahin H, Ciraci S. Stable, Single-Layer $MX_2$ Transition-Metal Oxides and Dichalcogenides in a Honeycomb-Like Structure. *J. Phys. Chem. C* **116**, 8983-8999 (2012).
23. Rasmussen FA, Thygesen KS. Computational 2D Materials Database: Electronic Structure of Transition-Metal Dichalcogenides and Oxides. *J. Phys. Chem. C* **119**, 13169-13183 (2015).
24. Inorganic Crystal Structure Database (ICSD). http://www2.fiz-karlsruhe.de/icsd_home.html
25. Gražulis S, *et al.* Crystallography Open Database (COD): an open-access collection of crystal structures and platform for world-wide collaboration. *Nucleic Acids Res.* **40**, D420-D427 (2012).
26. Ashton M, Paul J, Sinnott SB, Hennig RG. Topology-Scaling Identification of Layered Solids and Stable Exfoliated 2D Materials. *Phys. Rev. Lett.* **118**, 106101 (2017).
27. Choudhary K, Kalish I, Beams R, Tavazza F. High-throughput Identification and Characterization of Two-dimensional Materials using Density functional theory. *Sci. Rep.* **7**, 5179 (2017).
28. Mounet N, *et al.* Two-dimensional materials from high-throughput computational exfoliation of experimentally known compounds. *Nat. Nanotechnol.* **13**, 246-252 (2018).
29. Cheon G, Duerloo K-AN, Sendek AD, Porter C, Chen Y, Reed EJ. Data Mining for New Two- and One-Dimensional Weakly Bonded Solids and Lattice-Commensurate Heterostructures. *Nano Lett.* **17**, 1915-1923 (2017).
30. Junki S, Tsuyoshi Y, Yuki A, Kan N, Hiroyuki H. Epitaxial growth of silicene on ultra-thin Ag(111) films. *New J. Phys.* **16**, 095004 (2014).
31. Kappera R, *et al.* Phase-engineered low-resistance contacts for ultrathin $MoS_2$ transistors. *Nat. Mater.* **13**, 1128 (2014).

32. Sten H*, et al.* The Computational 2D Materials Database: high-throughput modeling and discovery of atomically thin crystals. *2D Mater.* **5**, 042002 (2018).
33. Doudin N*, et al.* Nanoscale Domain Structure and Defects in a 2-D $WO_3$ Layer on Pd(100). *J. Phys. Chem. C* **120**, 28682-28693 (2016).
34. Huang B*, et al.* Layer-dependent ferromagnetism in a van der Waals crystal down to the monolayer limit. *Nature* **546**, 270 (2017).
35. Cordero B*, et al.* Covalent radii revisited. *Dalton Trans.* , 2832-2838 (2008).
36. Ong SP*, et al.* Python Materials Genomics (pymatgen): A robust, open-source python library for materials analysis. *Comput. Mater. Sci.* **68**, 314-319 (2013).
37. Mathew K*, et al.* Atomate: A high-level interface to generate, execute, and analyze computational materials science workflows. *Comput. Mater. Sci.* **139**, 140-152 (2017).
38. Jain A*, et al.* FireWorks: a dynamic workflow system designed for high-throughput applications. *Concurr, Comp-Pract. E.* **27**, 5037-5059 (2015).
39. Kresse G, Hafner J. *Ab initio* molecular dynamics for liquid metals. *Phys. Rev. B* **47**, 558-561 (1993).
40. Kresse G, Hafner J. *Ab initio* molecular-dynamics simulation of the liquid-metal-amorphous-semiconductor transition in germanium. *Phys. Rev. B* **49**, 14251-14269 (1994).
41. Klimeš J, Bowler DR, Michaelides A. Chemical accuracy for the van der Waals density functional. *J. Phys.: Condens. Matter* **22**, 022201 (2009).
42. Klimeš J, Bowler DR, Michaelides A. Van der Waals density functionals applied to solids. *Phys. Rev. B* **83**, 195131 (2011).
43. Dion M, Rydberg H, Schröder E, Langreth DC, Lundqvist BI. Van der Waals density functional for general geometries. *Phys. Rev. Lett.* **92**, 246401 (2004).
44. Román-Pérez G, Soler JM. Efficient implementation of a van der Waals density functional: application to double-wall carbon nanotubes. *Phys. Rev. Lett.* **103**, 096102 (2009).
45. Björkman T, Gulans A, Krasheninnikov AV, Nieminen RM. van der Waals Bonding in Layered Compounds from Advanced Density-Functional First-Principles Calculations. *Phys. Rev. Lett.* **108**, 235502 (2012).
46. Setyawan W, Curtarolo S. High-throughput electronic band structure calculations: Challenges and tools. *Comput. Mater. Sci.* **49**, 299-312 (2010).
47. Ong SP, Wang L, Kang B, Ceder G. Li−Fe−P−O2 Phase Diagram from First Principles Calculations. *Chem. Mater.* **20**, 1798-1807 (2008).
48. Cai L*, et al.* Rapid Flame Synthesis of Atomically Thin $MoO_3$ down to Monolayer Thickness for Effective Hole Doping of $WSe_2$. *Nano Lett.* **17**, 3854-3861 (2017).
49. Chou SS*, et al.* Understanding catalysis in a multiphasic two-dimensional transition metal dichalcogenide. *Nat. Commun.* **6**, 8311 (2015).
50. Dávila ME, Xian L, Cahangirov S, Rubio A, Lay GL. Germanene: a novel two-dimensional germanium allotrope akin to graphene and silicene. *New J. Phys.* **16**, 095002 (2014).
51. Druffel DL*, et al.* Experimental Demonstration of an Electride as a 2D Material. *J. Am. Chem. Soc.* **138**, 16089-16094 (2016).
52. Xu S*, et al.* van der Waals Epitaxial Growth of Atomically Thin $Bi_2Se_3$ and Thickness-Dependent Topological Phase Transition. *Nano Lett.* **15**, 2645-2651 (2015).
53. Kong D*, et al.* Few-Layer Nanoplates of $Bi_2Se_3$ and $Bi_2Te_3$ with Highly Tunable Chemical Potential. *Nano Lett.* **10**, 2245-2250 (2010).
54. Li J*, et al.* Synthesis of 2D Layered $BiI_3$ Nanoplates, $BiI_3/WSe_2$ van der Waals Heterostructures and Their Electronic, Optoelectronic Properties. *Small* **13**, 1701034 (2017).
55. Pacilé D, Meyer JC, Girit ÇÖ, Zettl A. The two-dimensional phase of boron nitride: Few-atomic-layer sheets and suspended membranes. *Appl. Phys. Lett.* **92**, 133107 (2008).
56. Zhao H*, et al.* Fabrication of atomic single layer graphitic-$C_3N_4$ and its high performance of photocatalytic disinfection under visible light irradiation. *Appl. Catal., B* **152-153**, 46-50 (2014).

57. Elias DC, *et al.* Control of Graphene's Properties by Reversible Hydrogenation: Evidence for Graphane. *Science* **323**, 610-613 (2009).
58. Gong C, *et al.* Discovery of intrinsic ferromagnetism in two-dimensional van der Waals crystals. *Nature* **546**, 265 (2017).
59. Pei QL, *et al.* Spin dynamics, electronic, and thermal transport properties of two-dimensional $CrPS_4$ single crystal. *J. Appl. Phys.* **119**, 043902 (2016).
60. Lin M-W, *et al.* Ultrathin nanosheets of $CrSiTe_3$: a semiconducting two-dimensional ferromagnetic material. *Journal of Materials Chemistry C* **4**, 315-322 (2016).
61. Pozo-Zamudio OD, *et al.* Photoluminescence of two-dimensional GaTe and GaSe films. *2D Mater.* **2**, 035010 (2015).
62. Bianco E, Butler S, Jiang S, Restrepo OD, Windl W, Goldberger JE. Stability and Exfoliation of Germanane: A Germanium Graphane Analogue. *ACS Nano* **7**, 4414-4421 (2013).
63. Novoselov KS, *et al.* Electric Field Effect in Atomically Thin Carbon Films. *Science* **306**, 666-669 (2004).
64. 2D MaterialsXu K, *et al.* Ultrasensitive Phototransistors Based on Few-Layered $HfS_2$. *Adv. Mater.* **27**, 7881-7887 (2015).
65. Yue R, *et al.* $HfSe_2$ Thin Films: 2D Transition Metal Dichalcogenides Grown by Molecular Beam Epitaxy. *ACS Nano* **9**, 474-480 (2015).
66. Wang Y-Q, *et al.* Tunable Electronic Structures in Wrinkled 2D Transition-Metal-Trichalcogenide (TMT) $HfTe_3$ Films. *Adv. Electron. Mater,* **2**, 1600324 (2016).
67. Jiadong Z, *et al.* InSe monolayer: synthesis, structure and ultra-high second-harmonic generation. *2D Mater.* **5**, 025019 (2018).
68. Radisavljevic B, Radenovic A, Brivio J, Giacometti V, Kis A. Single-layer $MoS_2$ transistors. *Nat. Nanotechnol.* **6**, 147 (2011).
69. Tonndorf P, *et al.* Photoluminescence emission and Raman response of monolayer $MoS_2$, $MoSe_2$, and $WSe_2$. *Opt. Express* **21**, 4908-4916 (2013).
70. Wang Y, *et al.* Structural phase transition in monolayer $MoTe_2$ driven by electrostatic doping. *Nature* **550**, 487 (2017).
71. Du K-z, *et al.* Weak Van der Waals Stacking, Wide-Range Band Gap, and Raman Study on Ultrathin Layers of Metal Phosphorus Trichalcogenides. *ACS Nano* **10**, 1738-1743 (2016).
72. Wang X, *et al.* Chemical vapor deposition of trigonal prismatic $NbS_2$ monolayers and 3R-polytype few-layers. *Nanoscale* **9**, 16607-16611 (2017).
73. Xi X, *et al.* Strongly enhanced charge-density-wave order in monolayer $NbSe_2$. *Nat. Nanotechnol.* **10**, 765 (2015).
74. Shao Y, *et al.* Epitaxial fabrication of two-dimensional $NiSe_2$ on Ni(111) substrate. *Appl. Phys. Lett.* **111**, 113107 (2017).
75. Li L, *et al.* Black phosphorus field-effect transistors. *Nat. Nanotechnol.* **9**, 372 (2014).
76. Oyedele AD, *et al.* PdSe2: Pentagonal Two-Dimensional Layers with High Air Stability for Electronics. *J. Am. Chem. Soc.* **139**, 14090-14097 (2017).
77. Zhao Y, *et al.* Extraordinarily Strong Interlayer Interaction in 2D Layered $PtS_2$. *Adv. Mater.* **28**, 2399-2407 (2016).
78. Wang Y, *et al.* Monolayer $PtSe_2$, a New Semiconducting Transition-Metal-Dichalcogenide, Epitaxially Grown by Direct Selenization of Pt. *Nano Lett.* **15**, 4013-4018 (2015).
79. Tongay S, *et al.* Monolayer behaviour in bulk $ReS_2$ due to electronic and vibrational decoupling. *Nat. Commun.* **5**, 3252 (2014).
80. Yang S, *et al.* Layer-dependent electrical and optoelectronic responses of $ReSe_2$ nanosheet transistors. *Nanoscale* **6**, 7226-7231 (2014).
81. Weber D, Schoop LM, Duppel V, Lippmann JM, Nuss J, Lotsch BV. Magnetic Properties of Restacked 2D Spin 1/2 honeycomb $RuCl_3$ Nanosheets. *Nano Lett.* **16**, 3578-3584 (2016).
82. Saji KJ, Tian K, Snure M, Tiwari A. 2D Tin Monoxide—An Unexplored p-Type van der Waals Semiconductor: Material Characteristics and Field Effect Transistors. *Adv. Electron. Mater,* **2**, 1500453 (2016).

83. Liu G, *et al.* Vertically aligned two-dimensional $SnS_2$ nanosheets with a strong photon capturing capability for efficient photoelectrochemical water splitting. *Journal of Materials Chemistry A* **5**, 1989-1995 (2017).
84. Li L, *et al.* Single-Layer Single-Crystalline SnSe Nanosheets. *J. Am. Chem. Soc.* **135**, 1213-1216 (2013).
85. Tan C, *et al.* High-Yield Exfoliation of Ultrathin Two-Dimensional Ternary Chalcogenide Nanosheets for Highly Sensitive and Selective Fluorescence DNA Sensors. *J. Am. Chem. Soc.* **137**, 10430-10436 (2015).
86. Li L, *et al.* Ternary $Ta_2NiSe_5$ Flakes for a High-Performance Infrared Photodetector. *Adv. Funct. Mater.* **26**, 8281-8289 (2016).
87. Ryu H, *et al.* Persistent Charge-Density-Wave Order in Single-Layer $TaSe_2$. *Nano Lett.* **18**, 689-694 (2018).
88. Wan C, *et al.* Flexible n-type thermoelectric materials by organic intercalation of layered transition metal dichalcogenide $TiS_2$. *Nat. Mater.* **14**, 622 (2015).
89. Rui X, *et al.* Ultrathin $V_2O_5$ nanosheet cathodes: realizing ultrafast reversible lithium storage. *Nanoscale* **5**, 556-560 (2013).
90. Okada M, *et al.* Direct Chemical Vapor Deposition Growth of $WS_2$ Atomic Layers on Hexagonal Boron Nitride. *ACS Nano* **8**, 8273-8277 (2014).
91. Fei Z, *et al.* Edge conduction in monolayer $WTe_2$. *Nat. Phys.* **13**, 677 (2017).
92. Zhang M, *et al.* Controlled Synthesis of $ZrS_2$ Monolayer and Few Layers on Hexagonal Boron Nitride. *J. Am. Chem. Soc.* **137**, 7051-7054 (2015).
93. Mañas-Valero S, García-López V, Cantarero A, Galbiati M. Raman Spectra of $ZrS_2$ and $ZrSe_2$ from Bulk to Atomically Thin Layers. *Appl. Sci.* **6**, 264 (2016).
94. Tsipas P, *et al.* Massless Dirac Fermions in $ZrTe_2$ Semimetal Grown on InAs(111) by van der Waals Epitaxy. *ACS Nano* **12**, 1696-1703 (2018).